\documentclass[11pt]{article}

\usepackage{fullpage}
\usepackage{amsmath,amsfonts,amsthm,mathrsfs,mathpazo,xspace,hyperref,graphicx}
\usepackage{endnotes}
\usepackage{color}
\usepackage{bm}
\usepackage{times}
\usepackage{amssymb,latexsym}
\usepackage{paralist}

\usepackage{algorithm}
\floatname{algorithm}{Protocol}
\usepackage{algcompatible}

\newtheorem{theorem}{Theorem}

\newtheorem{lemma}[theorem]{Lemma}

\theoremstyle{remark}

\theoremstyle{definition}
\newtheorem{definition}[theorem]{Definition}

\newcommand{\beq}{\begin{eqnarray}}
\newcommand{\eeq}{\end{eqnarray}}

\newcommand{\eps}{\varepsilon}

\newcommand{\Hmin}{H_\infty}

\newcommand{\MS}{\textsc{MS}}

\newif\ifnotes\notesfalse

\begin{document}

\title{Parallel DIQKD from parallel repetition}

\author{Thomas Vidick\thanks{Department of Computing and Mathematical Sciences,
    California Institute of Technology, Pasadena, USA. email:
    \texttt{vidick@cms.caltech.edu}. Research funded by NSF CAREER Grant CCF-1553477, AFOSR YIP award number FA9550-16-1-0495, and the IQIM, an NSF Physics Frontiers Center (NFS Grant PHY-1125565) with support of the Gordon and Betty Moore Foundation (GBMF-12500028).} }
\date{}
\maketitle


\vspace{-1cm}

\begin{abstract}
We give an arguably simpler and more direct proof of a recent result by Miller, Jain and Shi, who proved device-independent security of a protocol for quantum key distribution in which the devices can be used in parallel. Our proof combines  existing results on immunization (Kempe et al., SICOMP 2011) and parallel repetition (Bavarian et al., STOC 2017) of entangled games. 
\end{abstract}

In a recent preprint~\cite{miller2017parallelQKD}, Miller et al. give a protocol for device-independent quantum key distribution (DIQKD) in which the users provide inputs to, and collect outpus from, their respective devices \emph{in parallel}: Alice (resp. Bob) selects a random string of $N$ inputs ${ x}=x_1,\ldots,x_N\in \mathcal{X}$ (resp. ${y}= y_1,\ldots,y_N\in \mathcal{Y}$); each user provides its $N$ inputs to its respective device and collects $N$ outputs ${a}=a_1\ldots,a_N\in\mathcal{A}$ (resp. ${ b} = b_1,\ldots,b_N\in\mathcal{B}$. Once this phase has completed the devices are no longer needed. The protocol concludes by classical phases of parameter estimation, error correction and privacy amplification.  

The proof in~\cite{miller2017parallelQKD} introduces a number of novel techniques in order to analyze the entropy generation, as well as the robustness, of the protocol, which is based on the Mermin-Peres Magic Square game~\cite{Arvind:02} as a certificate of entropy generation. The goal of this note is to sketch a different proof of the same result, obtained by an elementary combination of existing results. The first result is the technique of ``immunization'' introduced in~\cite{KKMTV11}: this technique provides a generic method to show that a three-player guessing game based on (e.g.) the Magic Square game cannot be won with probability $1$, even by players sharing entanglement; see Lemma~\ref{lem:imm}. The second result is a threshold theorem for the parallel repetition of multiplayer entangled games that satisfy a property called ``anchored''; see Lemma~\ref{lem:threshold}. Combining these two results gives a proof of security of a similar (though subtly different) protocol for parallel DIQKD than the one in ~\cite{bavarian2015anchoring}; see Section~\ref{sec:protocol}. 

In this note we sketch the simple argument, hoping to provide an alternative viewpoint on~\cite{miller2017parallelQKD}. We omit the more standard details, and do not comment on the usefulness or practicality of parallel DIQKD. 

\section{Notation}

For a string $x\in \mathcal{X}^n$ and $S\subseteq\{1,\ldots,n\}$ we let $x_S$ be the bits of $x$ indexed by $S$. 
Given a multi-player game $G$, we let $\omega_c(G)$ and $\omega^*(G)$ be its classical and entangled value respectively. 

We use $\MS$ to denote the Magic Square game, which is such that $\omega_c(\MS)=1$ and $\omega^*(\MS)=1$. The Magic Square game is a free game (i.e. the input distribution has a product form) in which each player has three possible inputs $x\in\mathcal{X}$ (a row), $y\in\mathcal{Y}$ (a column) and four possible outputs $a\in\mathcal{A}$, $b\in\mathcal{B}$ (an even or odd assignment to the entries in the row or column). It has the useful property that for every $(x,y)\in\mathcal{X}\times\mathcal{Y}$ there exists functions 
\begin{equation}\label{eq:def-fg}
f_{xy}:\mathcal{A}\to\{0,1\},\qquad g_{xy}:\mathcal{B}\to\{0,1\}
\end{equation}
 such that for any valid output-input tuple $(a,b|x,y)$ in the game, $f_{xy}(a)=g_{xy}(b)$. In other words, there is always one bit that is expected to match in each players' answers. 


\section{Guessing games}

\begin{definition}\label{def:guessing}
Let $G$ be a two-player free game, and $0<\eta \leq 1$. We define the $\eta$-guessing game associated with $G$, $G_{\eta}$, as follows: 
\begin{compactenum}
\item Alice and Bob receive independent inputs $x,y$ respectively, distributed as in $G$. 
\item With probability $(1-\eta)$ Eve receives input $(x,y)$. With probability $\eta$ she receives no input. 
\item The players produce outputs $a,b$ and $e$ respectively. 
\item The verifier accepts if and only if $(a,b|x,y)$ is a valid output-input tuple in $G$, and either $e=a$ or Eve had no input. 
\end{compactenum}
\end{definition}

The following lemma follows from the ``immunization'' technique introduced in~\cite{KKMTV11} (see e.g. Lemma~17 in the paper). 

\begin{lemma}\label{lem:imm}
Let $G$ be a two-player game such that $\omega^*(G) = 1  > \omega_c(G)$. Then for any $0<\eta \leq 1$ there is a $C_G(\eta) > 0 $ (depending on $\eta$ and the number of questions in $G$) such that  
 $$\omega_c(G) \,\leq\, \omega^*(G_\eta) \,\leq\, 1 - C_G(\eta).$$  
\end{lemma}

Although our results apply more generally, to fix ideas we focus on a game $G$ instantiated as the Magic Square game $\MS$, and $\eta = 1/8$ (this is an arbitrary choice). Furthermore, in Definition~\ref{def:guessing} we relax the requirement on Eve to only guess the bit $f_{xy}(a)=g_{xy}(b)$ in common in the players' answers (when they satisfy the winning condition for $\MS$). It can be shown using the same immunization technique that Lemma~\ref{lem:imm} still holds with this requirement (see also Proposition 4.1 in~\cite{miller2017parallelQKD}). Let $C_\MS^*  = C_\MS(1/8) > 0$ be the constant associated to this choice of game and $\eta$ by Lemma~\ref{lem:imm}, i.e.
\begin{equation}\label{eq:def-omegams}
C_\MS^* \,=\, 1-\omega^*(\MS_{1/8}).
\end{equation}

We now consider the problem of parallel repetition of a multiplayer game.

\begin{definition}\label{def:threshold-game}
Let $G$ be a multiplayer game, $n\geq 1$ an integer and $\omega^*(G)\leq t \leq 1$ a threshold value. We define $\tau^*_{n,t}(G)$ to be the entangled value of the following game $G^{(n,t)}$:
\vskip.1cm 
\begin{compactitem}
\item The referee selects $n$ independent tuples of inputs for the players in $G$, and simultaneously sends each player its $n$ respective inputs; each player replies with $n$ outputs.
\item The referee accepts if and only if the fraction of rounds in which the winning condition for $G$ is satisfied by the players' inputs and outputs for that round is at least $t$. 
\end{compactitem}
\end{definition}

The following follows from~\cite[Theorem 23]{bavarian2015anchoring}. The only condition to verify is that for any two-player free game $G$ and $\eta > 0$ the game $G_\eta$ is an anchored game, which is immediate from the definition (this is the sole reason for introducing $G_\eta$ from $G$).  

\begin{lemma}\label{lem:threshold}
Let $G$ be a two-player free game, $0<\eta \leq 1$ and $\delta>0$ such that $t = \omega^*(G_\eta)+\delta \leq 1$. Then 
$$ \tau_{n,t}^*(G) \,\leq\, e^{-\Omega(\delta^9 n)},$$ 
where the implicit constant in the exponent depends on $\eta$ and $|G|$ but not on $n$. 
\end{lemma}

\section{Parallelizing DIQKD}
\label{sec:protocol}

We consider a simple protocol for parallel DIQKD, Protocol~\ref{pro:parallel}, directly inspired from the protocol MagicQKD in~\cite{miller2017parallelQKD}. The following theorem states a bound on the quantum conditional min-entropy of Alice's outputs at the end of the protocol. Applying standard steps of error correction and privacy amplification it is straightforward to obtain a positive key rate from the theorem. (Using that the Magic Square game has the property that in a winning strategy one of Alice's output bits is required to equal one of Bob's output bits, the additional loss due to error correction will scale as $O(\eps n)$.)

\begin{theorem}
Let $\MS$ be the two-player Magic Square game, $C^*_{\MS}>0$ the constant defined in~\eqref{eq:def-omegams}, and $\gamma,\eps >0$ such that $\eps < C^*_{\MS}/2$. Suppose that Protocol~\ref{pro:parallel} (with parameter $\eta=1/8$) is executed with arbitrary devices such that the probability of Alice and Bob aborting in Step~\ref{step:check} is at most $p_a$. 

 Let $\rho_{K_A E}$ be the joint state of Alice's raw key and Eve's side information at the end of the protocol, conditioned on Alice and Bob not aborting in Step~\ref{step:check}. 
Then 
$$\Hmin^{\eps_s}(K_A | E)_{\rho} \geq \Omega( (C^*_\MS-2\eps)^9 n) -\log p_a^{-1} - O(\gamma n),$$
where $\eps_s = p_a^{-1}\exp(-\Omega(\eps^2 \gamma n))$. Moreover, honest players using $(\eps/2)$-noisy devices are accepted in the protocol with probability $1-\exp(-\Omega(\eps^2\gamma)n)$. 
\end{theorem}

The bound claimed in the theorem is analogous to~\cite[Theorem 1.2]{miller2017parallelQKD}. We do not work out explicit constants, but due to the protocol being simpler, and the analysis more direct, we expect that they could be made to improve upon~\cite{miller2017parallelQKD}. 

\begin{proof}
Let $G = \MS$ and $\eta = 1/8$. Observe that right after Step~\ref{step:shareinputs} in Protocol~\ref{pro:parallel} the inputs in the possession of Alice, Bob and Eve are distributed exactly as in the $n$-fold parallel repetition of the game $G_\eta$: Alice and Bob have $n$ independent inputs to $G$, while Eve has both player's inputs in a subset $S$ of the rounds of expected size $(1-\eta)n$, and no input for the remaining rounds. 
Let $t=1-2\eps$. The winning condition for $G_\eta^{(n,t)}$ (Definition~\ref{def:threshold-game}) is implied by the conjunction of the following two conditions:
\vskip.1cm
\begin{compactitem}
\item Alice and Bob's outputs satisfy the winning condition for $G$ in a fraction at least $t$ of the rounds;
\item Eve's output $e_i$ matches $f_{x_iy_i}(a_i)$ in all rounds $i\in S$.
\end{compactitem}
\vskip.1cm
We evaluate the probability that the first condition is not satisfied, yet the players do not abort at Step~\ref{step:check}. For $i\in\{1,\ldots,n\}$ let $W_i$ be the indicator random variable for the event that inputs and outputs for Alice and Bob in the $i$-th round satisfy the winning condition for $G$. Since the rounds $T$ in which the players evaluate the game condition are chosen uniformly, it follows from a standard concentration bound (see e.g.~\cite[Lemma 7]{tomamichel2015rigorous}; note that no independence is required of the $W_i$) that 
\begin{equation}\label{eq:conc}
 \Pr\Big( \sum_{i\in T} W_i > (1-\eps)|T| \wedge \sum_{i\in \{1,\ldots,n\}} W_i \leq (1-2\eps)n\Big)  =  e^{-\Omega(\eps^2 \gamma n)},
\end{equation}
where we may assume that the bound on the right-hand side incorporates the probability that Alice and Bob abort due to $|T| < \gamma n$, which given $\eta = 1/8$ and $\gamma \leq 1/2$ is exponentially small in $n$. 
Let $\tilde{\rho}_{K_AE}$ be the joint state of Alice's raw key and Eve's side information at the last step of the protocol, conditioned on the event that the players do not abort in Step~\ref{step:check}, and the condition $\sum_{i\in \{1,\ldots,n\}} W_i > (1-2\eps)n$ holds. Let $\tilde{p}_a$ be the probability of the latter conjunction of events. By definition of the winning condition for $G_\eta^{(n,t)}$ and the relation between guessing entropy and conditional min-entropy~\cite{konig2009operational} it follows that 
$$ \Hmin(K_A | E)_{\tilde{\rho}} \,\geq\, -\log \big(\tau^*_{n,t}(G)/\tilde{p}_a\big) .$$
Applying Lemma~\ref{lem:threshold}, $\tau^*_{n,t}(G) = \exp(-\Omega((t-(1-C^*_{\MS}))^9 n))$. Since by~\eqref{eq:conc} we have $\|\tilde{\rho} - \rho\|_1 = p_a^{-1} \exp(-\Omega(\eps^2\gamma n))$ (with $\rho = \rho_{K_AE}$ as defined in the theorem), we deduce the bound claimed in the theorem, where the subtraction of an $O(\gamma n)$ term accounts for outputs leaked to Eve in Step~\ref{step:check}.

Finally, the ``moreover'' part of the theorem follows from a standard concentration argument. 
\end{proof}

\begin{algorithm}[t]
\caption{Parallel DIQKD protocol}
\label{pro:parallel}
\begin{algorithmic}[1]
	\STATEx \textbf{Arguments:} 
		\STATEx\hspace{\algorithmicindent} $D$ -- untrusted device 
		\STATEx\hspace{\algorithmicindent} $n \in \mathbb{N}_+$ -- number of rounds
		\STATEx\hspace{\algorithmicindent} $\eta \in [0,1)$ -- fraction of rounds in which Alice and Bob's inputs are not leaked to Eve (\emph{game} rounds).
		\STATEx\hspace{\algorithmicindent} $\gamma \in (0,1/2]$ -- fraction of rounds in which Alice and Bob test the game condition (\emph{test} rounds). 
		\STATEx\hspace{\algorithmicindent} $\eps\in[0,1/2]$ -- noise tolerance for honest devices.  
	\STATEx
	\STATE For every $i\in\{1,\ldots,n\}$, Alice and Bob independently select inputs $x_i$ and $y_i$ in the game $G$. 
	\STATE\label{step:shareinputs} Alice selects a random subset $S \subseteq \{1,\ldots,n\}$ by choosing each round independently with probability $(1-\eta)$. She sends $(S,x_S)$ to Bob. Bob replies with $y_S$. 
	\STATE Alice and Bob provide their respective strings of inputs, $x$ and $y$, to their device.
	\STATE Alice and Bob collect output strings $a$ and $b$ from their respective device. 
\STATE\label{step:check} Alice selects a random subset $T\subseteq S$ of size $|T| = \gamma n$ (if $|S|\leq \gamma n$ they abort). She sends $(T,a_T)$ to Bob. Bob replies with $b_T$. They abort the protocol if fewer than $(1-\eps)|T|$ of the rounds in $T$ satisfy the winning condition for $G$. 
 \STATE Alice (resp. Bob) sets $(K_A)_i = f_{x_iy_i}(a_i)$ (resp. $(K_B)_i = g_{x_iy_i}(a_i)$, for $i\in S$, where $f,g$ are as in~\eqref{eq:def-fg}. The resulting $S$-bit strings form their raw key. 
	\end{algorithmic}
\end{algorithm}

\bibliography{../main}


\end{document}